# Large-Gap Quantum Spin Hall State and Temperature-Induced Lifshitz Transition in Bi$_4$Br$_4$

Ming Yang[1], Yundan Liu[2], Wei Zhou[3], Chen Liu[4,*], Dan Mu[2], Yani Liu[5], Jiaou Wang[4], Weichang Hao[1], Jin Li[2,*], Jianxin Zhong[2], Yi Du[1,5], Jincheng Zhuang[1,*]

[1]*School of Physics and BUAA-UOW Joint Research Centre, Beihang University, Haidian District, Beijing 100191, China*

[2] *Hunan Key Laboratory of Micro-Nano Energy Materials and Devices, and School of Physics and Optoelectronics, Xiangtan University, Hunan 411105, China*

[3] *School of Electronic and Information Engineering, Changshu Institute of Technology, Changshu, 215500, China*

[4] *Beijing Synchrotron Radiation Facility, Institute of High Energy Physics, Chinese Academy of Sciences, Beijing 100049, P. R. China*

[5] *Institute for Superconducting and Electronic Materials, Australian Institute for Innovative Materials, University of Wollongong, Wollongong, NSW 2500, Australia*

Searching for new quantum spin Hall insulators with large fully opened energy gap to overcome the thermal disturbance at room temperature has attracted tremendous attention due to the one-dimensional (1D) spin-momentum locked topological edge states serving as dissipationless channels for the practical applications in low consumption electronics and high performance spintronics. Here, we report the investigation of topological nature of monolayer Bi$_4$Br$_4$ by the techniques of scanning tunneling microscopy and angle-resolved photoemission spectroscopy (ARPES). The topological non-triviality of 1D edge state integrals within the large bulk energy gap (~ 0.2 eV) is revealed by the first-principle calculations. The ARPES measurements at different temperature show a temperature-induced Lifshitz transition, corresponding to the resistivity anomaly caused by the shift of chemical potential. The connection between the emergency of superconductivity and the Lifshitz transition is discussed.

Quantum spin Hall (QSH) insulators are the two-dimensional (2D) topological materials that host one-dimensional (1D) conducting edge states bridging the 2D bulk band gap [1-3]. The electrons at the topological non-trivial edge states possess the locked spin and momentum, *i.e.*, different spins when counter-propagating, prohibiting the only scattering channel, back scattering, in 1D helical edge current from the nonmagnetic impurities as protected by time reversal symmetry [4-7]. These features make QSH insulators the promising candidate for the applications of the spin-based spintronics and quantum computers. Since the prediction for graphene and the experimental realization in HgTe quantum wells [3,4] of the QSH effect [1,2,4,8], tremendous efforts have been made for searching new 2D topological insulators and advancing the understanding of this novel quantum state. Nevertheless, most of the experimentally confirmed candidates of QSH insulators are impractical for applications due to either the small bandgaps or strong interaction with the underlying substrates. The size of the QSH bandgap, which is primarily determined by the strength of spin-orbit coupling (SOC), must be substantially larger than the thermal broadening at the operating temperature to promise the chemical potential residing within the gap and to suppress thermal excitation of charge carriers of edge states into the bulk electron bands [9,10]. A gap magnitude significantly larger than 0.1 eV is required for room-temperature operation according to the formula $E_g \gg 3.5\ k_B T$, where $E_g$ and $k_B$ are the bandgap and Boltzmann constant, respectively. QSH state with large gap compatible to practical level has been identified in monolayer Na$_3$Bi on Si(111) [11], stanene on Cu(111) [12], and bismuthene on SiC(0001) [13]. Nevertheless, the 3D out-of-plane bonding between these QSH insulators and the substrates makes the detrimental contributions to the further practical manufacture, such as mechanical isolation of monolayer QSH insulators and realization of nanoribbons by lithographic patterning technique. Furthermore, the sizes of 1D topological edges are constricted by quality and shape of 2D QSH insulators, hindering the formation of long-straight nanoribbons, which hold the potential for nanoscale electronic application due to their small conducting channel sizes. Thus, the large-gap QSH insulators with atomically sharp and long-straight edges are highly desired for widespread study and applications.

The Bismuth halide, Bi$_4$X$_4$ ($X$ = I or Br), hold the structural primitive of 1D infinite molecular chain, where both of their 2D and three-dimensional (3D) allotropes are stacked through van der Waals force. A variety of topological phases, including weak topological insulator (WTI) and high-order topological insulator (HOTI), have been experimentally identified in their 3D allotropes resulted from different out-of-plane stacking orders [14-17]. For 2D allotropes in this system, the first-principle calculations predict a large-gap QSH insulator and Dirac semimetal in monolayer Bi$_4$Br$_4$ and Bi$_4$I$_4$, respectively, where the topological phase transition between trivial insulator and QSH insulator could be effectively tuned by introducing the symmetry-breaking perturbations as the monolayer Bi$_4$I$_4$ is near the boundary of this phase transition [18]. The Bi$_4$Br$_4$ 1D molecular chain is an ideal platform to engender topological nanoribbons with clean and atomically sharp edges. The QSH state of monolayer Bi$_4$Br$_4$, however, has not been experimentally demonstrated due to the absence of the high-spatial-resolution characterization of local electronic properties of 2D Bi$_4$Br$_4$ surface and its 1D edges. Furthermore, the transport measurements with a resistivity peak as a function of temperature, which may be correlated to the structural phase transition, charge/spin density wave, semiconductor-semimetal transition or temperature-induced Lifshitz transition in other topological material, exhibit unclear mechanism [19-23]. Direct investigation of the surface electronic properties of Bi$_4$Br$_4$ is highly demanded to reveal its topological nature and origin of transport property anomaly.

In this work, we successfully synthesized the high-quality Bi$_4$Br$_4$ single crystal by the solid-state reaction method, similar to the ref[17]. High-purity Bi and BiBr$_3$ powder with the molar ration 1:1 was ground thoroughly and then were sealed in quartz tube under vacuum. The sealed raw materials were placed in two-heating-zone furnace with a temperature gradient from 558 K to 461 K and kept at these temperatures for 72 h. After cooling down, the single crystal Bi$_4$Br$_4$ nucleates at the hot side of the quartz tube. The typical morphology of single crystal Bi4Br4 is needle-liked shape with the size around $5.0 \times 0.5 \times 0.2$ mm$^3$.

The large QSH bandgap and the 1D topological non-trivial edge state have been identified in the top-layer Bi$_4$Br$_4$ surface by the combined techniques of scanning tunneling microscopy (STM) and angle-resolved photoemission spectroscopy (ARPES). The STM measurements were carried out using a low-temperature UHV STM/scanning near-field optical microscopy system (SNOM1400, Unisoku Co.), where the bias voltages were applied to the substrate. The differential conductance, dI/dV, spectra were acquired by using a standard lock-in technique with a 10 mV modulation at 973 Hz. All the STM measurements were performed at 77 K. The ARPES characterizations were performed at the Photoelectron Spectroscopy Station in the Beijing Synchrotron Radiation Facility using a SCIENTA R4000 analyzer. A monochromatized He I light source (21.2 eV) was used for the band dispersion measurements. The total energy resolution was better than 15 meV, and the angular resolution was set to ~0.3°, which gives a momentum resolution of ~0.01 π/a. The high-pressure electrical transport properties of Bi$_4$Br$_4$ crystal were measured through a diamond anvil cell (DAC) made of Cu-Be alloy. Sodium chloride was applied as the transmitting medium. The pressure was determined by monitoring the shift of the fluorescence peak of a ruby sphere at room temperature.

The density functional theory (DFT) calculations reveal the non-trivial topological invariant and edge state in Bi$_4$Br$_4$ nanoribbons after considering the van der Waals interaction between the nanoribbons and underlying bulk Bi$_4$Br$_4$. Our first-principles calculations were performed by using the Vienna ab initio simulation package (VASP) [24] with the projector augmented wave approach [25] and the Perdew–Burke–Ernzerh generalized-gradient approximation [26]. The cutoff energy of the plane-wave expansion of the basis functions was set to be 450 eV and the Brillouin-zone integrations were performed using 3 × 9 × 1 and 1 × 9 × 1 for 2D structures and nanoribbons, respectively. Due to weak van der Waals interactions in Bi$_4$Br$_4$, the vdW corrections of Grimme DFT-D3 method [27] were employed in our calculations. The convergence criteria for electronic and ionic relaxations are $10^{-5}$ eV and $10^{-3}$ eV/Å, respectively. The vacuum thickness of the slab model was set to 20 Å to avoid spurious interactions between adjacent images. The $Z_2$ topological invariant was determined by calculating the evolution of the Wannier function center in reciprocal space during a "time-reversal pumping" process [28,29].

The temperature evolution of band structure indicates a marked down-shift trend with decreasing temperature, leading to an evolution from semiconductor above 180 K to *n*-type semimetal with an electron-like Fermi pocket at low temperature. The results are consistent with resistivity anomaly, and provide direct electronic evidence of a temperature-induced Lifshitz transition. Our observations indicate the QSH insulating nature and the relationship between the Lifshitz transition and the pressure-induced superconductivity of Bi$_4$Br$_4$.

The photographic images of the Bi$_4$Br$_4$ crystals we applied in our experiments are displayed in the inset of FIG. 1(a) and FIG. S1. The samples are metal needles with the length up to 7 mm and the width up to 1 mm, respectively. X-ray diffraction measurement displayed in FIG. 1(a) confirmed a bilayer structure in the crystal along the *c* axis. In fact, previous studies reported that the structure of bulk crystals is *α*' bilayer phase [17,30], as shown in FIG. 1(e). The building block of *α*' Bi$_4$Br$_4$ is the 1D molecular chains along the *b* axis with four Bi atoms in width and terminated by Br atoms. The Bi atoms could be divided into two types, where the inner Bi atom (Bi$_{in}$) is surrounded by three nearest neighboring Bi atoms, and the external Bi atom (Bi$_{ex}$) is bind to four Br atoms. Along the *c* axis, the adjacent layers are shifted with respect to each other by half of the chain translation, *i.e.*, *b*/2 lattice, and flipped by 180 degrees, forming the so-called AB-stacking model.

The large-scale STM topography of cleaved Bi$_4$Br$_4$ surface is displayed in FIG. 1(b) and FIG. S2, where long-straight nanoribbons with the length up to micrometer have been deduced. Furthermore, the nanoribbons hold the almost atomic sharp edges with the width discrepancy of only one or two chains. The formation of the ideal nanoribbons results from the unique van der Waals force of intralayer interaction. The prime step height is around 0.98 nm, as shown in FIG. 1(c), which is close to the out-of-plane height of single layer Bi$_4$Br$_4$ (10.03 × sin107.41° ~ 0.96 nm) [31,32]. FIG. 1(d) shows the high-resolution STM image of Bi$_4$Br$_4$ with two adjacent layers, where a mirror symmetry along the *b* axis could be identified by the directions of red arrow and white arrow for upper and lower layers, respectively. Such a mirror symmetry is caused by the 180 degrees rotation between adjacent layers of *α*' phase. The high-resolution STM images of Bi$_4$Br$_4$ terrace surface are displayed in FIG. 1(g) with the unit cell labelled by the red rectangle. The lattice constants along the short side and long side of this rectangle are 0.44 nm and 1.30 nm, consistent with previous reports and corresponding to *b* and *a* value [31,32], respectively. In order to figure out the detailed structures of the protrusions in STM topology, first-principles calculations were performed based on the monolayer Bi$_4$Br$_4$ atomistic model. The evolution from the atomic structure model to the simulated STM image and then to the experimental STM image in FIG. 1(h). is accordingly consistent with each other, indicating that brightest protrusion and the others in FIG. 1(g) are attributed to the top Bi atoms and Br atoms, respectively. Although the top Br atoms is close to upper surface compared to top Bi atoms from the view of the structural model in FIG. 1(e), the signals from both of the top Br atoms and Bi atoms are expected to be probed due to their small height difference (~ 0.76 Å) and the fact that STM images are contributed by surface density of states (DOS) of the sample. This deduction is confirmed by the calculated results of charge density difference in FIG. 1(f) and orbit-projected structures of monolayer $Bi_4Br_4$ in FIG. S3, where DOS of $Bi_4Br_4$ are made up by $p$ orbits of Bi atoms at the low energy near Fermi level ($E_F$).

FIG. 2(a-c) plots the ARPES intensities of the constant energy contours (CEC) at different binding energies of 0, 500, and 800 meV along the $k_y$-$k_x$ plane, where $k_y$ is correlated to the chain direction. The CEC at the $E_F$ displays an oval locating at the $M$ point of Brillouin zone (FIG. 2(a)). With the increment of the binding energy, it firstly shrinks and then grows in area. The oval at the Brillouin center, $\Gamma$ point, could be detected at the binding energy of 600 meV (Supplementary Note 4), and then overlaps and hybridizes with the signal from $M$ point to form a highly anisotropic band dispersion (FIG. 2(c)). This highly anisotropic band dispersion originates from the weak van der Waals intralayer interaction along the direction perpendicular to the chain, which is consistent with previous results [17]. FIG. 2(d) and 2(e) display the energy-momentum dispersion along the high-symmetry momentum cuts across the $\Gamma$ and $M$ points of Brillouin zone, respectively. The hole-like band at $\Gamma$ point locates at the energy level of 0.6 eV below the Fermi surface (FIG. 2(d)), and the low-energy bands are distributed near the $M$ point with a bandgap around 0.23 eV between electron-like pocket, upper band (UB), and hole-like pocket, lower band (LB). In view of the 1D band dispersion caused by the weakest bonding along the $c$-axis [17], the band structure measured along the $k_y$-$k_x$ plane could correspond to the electronic features of in-plane surface. Thus, the ARPES results reveal an energy gap between the UB and LB of $Bi_4Br_4$ surface at 6 K.

The surface electronic structures of $Bi_4Br_4$ were further studied by tunneling spectra d$I$/d$V$, which is proportional to the local DOS of the surface. The d$I$/d$V$ spectrum acquired in the terrace far away from the edge of $Bi_4Br_4$ represents a full open energy gap near $E_F$, as shown in FIG. 2(f), which is aligned in energy with the high-symmetry ARPES spectrum along the $\Gamma$-$M$-$\Gamma$ direction. An excellent agreement between STM and ARPES results on the position and the size of the energy gap further illustrates that the band structure obtained from ARPES measurements derives from the surface states. Compared to the gap feature in the terrace, the scanning tunneling spectroscopy (STS) spectrum detected at the edge position shows an obvious nonzero DOS filling in the gap (FIG. 2(g)), indicating the existence of a conductive edge state. FIG. 2(h) displays the spatially resolved d$I$/d$V$ spectra as a function of energy and distance across the edge in FIG. 2(i), demonstrating that the conductive state is confined at the edge of $Bi_4Br_4$ with a spatial extension close to the 1.7 nm, as labelled by the yellow and purple arrows. It should be noted that penetration length of the conductive edge state is comparable to that of the topological 1D edge state observed in other systems [33-38]. Nevertheless, the conductive edge state could be evoked by the crystal-symmetry-broken factors, such as the dangling bonds at the edge and various defects. In fact, the straight edges along the $b$-axis (stripe) direction of $Bi_4Br_4$ are essentially distinct from other cases from the view of the existence of the dangling bond as the inter-stripe interaction is attributed to van der Waals force, excluding the possibility of the dangling bond effect on the formation of conductive edge state. Additionally, there are several kinds of defects on the surface of single crystal $Bi_4Br_4$, as shown in FIG. S6. The spatially resolved STS data show that the bulk energy gap run continuously across the defect without any trace of existence of conductive edge state residing in the gap, ruling out the relationship between the defects and the edge state.

In order to figure out the topological nature of $Bi_4Br_4$ surface, we have studied their electronic properties by the first-principle calculations. The calculated band structure of monolayer $Bi_4Br_4$ without SOC, as shown in FIG. S4, shows a direct band gap located at $\Gamma$ point. From the view of projected DOS, both of the valence and conduction bands near the Fermi level are mainly contributed by Bi-6$p$ orbitals, and the states dominated by Br-4$p$ orbitals are pushed far away from Fermi surface due to the larger electronegativity of Br atoms. After the spin orbit coupling (SOC) is considered, the constituents and parities of the conduction and valence bands are inverted at $\Gamma$ point. The SOC-induced band inversion with the opposite parities indicates the non-trivial topological invariant in monolayer $Bi_4Br_4$. Based on the evolutions of Wannier function centers in Brillion zone [39,40], we have calculated the $Z_2$ topological invariant within the first-principle framework, where the $Z_2 = 1$ verifies that the monolayer $Bi_4Br_4$ belongs to the QSH insulators. More detailed information on the physical mechanism of the band inversion is listed in FIG. S7. After taking the edge potential and structure relaxation into account, we calculated the electronic structure of $Bi_4Br_4$ monolayer ribbon with the edge termination along the $b$ axis. The topological edge states (labelled by red lines) with Dirac-like shape could be clearly identified integrating within the energy gap of bulk electronic structure (black lines) in FIG. 3(a), which is consistent with the previous simulations [18]. The calculated energy gap of bulk electronic structure is around 0.18 eV, which agrees well with our ARPES results and STM curves. Since our sample is the $Bi_4Br_4$ single crystal with nanoribbons on the top of bulk $Bi_4Br_4$ surface, and the interface interaction may modulate the properties, the calculation of the electronic structure of $Bi_4Br_4$ nanoribbon with the support of monolayer $Bi_4Br_4$ are performed. The calculated results in FIG. 3(b) quantitatively resemble that of the freestanding monolayer $Bi_4Br_4$ ribbon, indicating that the weak interlayer interaction engenders the neglectable effect on the topological nature of the edge state. The penetration depth of topological edge state is 1.7 nm in FIG. 3(c) and (d), which is in accordance with the experimental findings in FIG. 2(h) and comparable to other 2D topological insulators [33-38]. It is noticeable that the topological trivial electronic edge state arising from edge dangling bonds or atomic reconstructions is highly restricted on the edge and decays exponentially away from the edge [41,42]. All the simulations reveal the topological non-trivial nature of monolayer $Bi_4Br_4$ and reproduce the experimental observations well.

Our low-temperature ARPES and STM results combined with the DFT simulations demonstrate the topological non-trivial nature of monolayer $Bi_4Br_4$ with a large gap. Nevertheless, the typical practical level of the application requires the operation at room temperature and above. Therefore, the temperature dependent ARPES measurements from low-temperature to room temperature need to be investigated to resolve the evolution of the electronic properties of $Bi_4Br_4$ and to examine on the possible temperature-induced topological phase transition. FIG. 4(a) displays the temperature dependence of the energy-momentum distribution measured crossing the $M$ points of Brillouin zone. The white dashed line is provided for the guidance of location of $E_F$, where the band dispersion above the $E_F$ could be revealed at the relatively high temperature due to the thermal excitation of electrons. At the low temperature (~ 6 K), the Fermi surface crosses the UB. The overall band structure shows a strong temperature dependence and shifts toward to the Fermi surface with increasing temperature. When the temperature reaches 180 K and above, the UB become invisible below $E_F$, and the Fermi surface enters into the gap between UB and LB. Similar band shift trend could also be observed on the ARPES data crossing the $\Gamma$ point of Brillouin zone (FIG. S8). These results show a clear temperature dependent variation of Fermi surface topology, which is correlated to the Lifshitz transition. The temperature-induced Lifshitz transition has been previously reported in the $WTe_2$ and $ZrTe_5$ [23,43]. Moreover, the Lifshitz transition temperature (~ 180 K) agrees well with that of the anomaly of transport properties of $Bi_4Br_4$, as shown in FIG. 4(b). At the high temperature, the Fermi level lying in the gapped region results in a semiconducting state of $Bi_4Br_4$, which corresponds to semiconducting behavior of temperature dependent resistivity, *i.e.*, resistivity increases with decreasing temperature. When the temperature downs to around 180 K, the UB touches the Fermi level and retains sinking down with the decrement of temperature, giving rise to *n*-type semimetal of $Bi_4Br_4$ as well as the reduced resistivity at the temperature below 170 K. Our temperature dependent ARPES data provides a natural explanation on the resistivity maximum at ~ 170 K of the transport measurement. The shift of the Fermi surface relative to the band structure is correlated to the variation of the chemical potential [23]. This temperature-induced chemical potential shift without the charge carrier doping is taken for the intrinsic properties of semiconductor, and has been identified in the degenerate semiconductor [23,44]. In fact, the high density of defects with several different types has been observed in our STM images (FIG. 1(b) and FIG. S3), which is expected to be the origin of temperature-induced chemical potential shift as well as the Lifshitz transition.

In previous reports, the Lifshitz transition-induced reconstruction of Fermi surfaces is correlated to variation of Fermi-level nesting conditions, formation of magnetic or charge ordering, and superconducting transitions in transition metal dichalcogenide, pnictides, and other related materials [23, 43, 45, 46]. The electrical resistance measurement for the $Bi_4Br_4$ single crystal is performed under quasi hydrostatic pressure. FIG. 4(b) shows the typical electrical resistance as a function of temperature measured under zero magnetic field in the pressures ranging from 0 to 5.4 GPa. Upon increasing pressure to 1.9 GPa, the resistance shows a semiconducting behavior in the whole temperature range, indicating the suppression of the Lifshitz transition to low temperature. Intriguingly, an abrupt drop of resistance at 6 K coexists with the suppression of Lifshitz transition under this pressure. A sharp drop to zero resistance could be identified at higher pressure (~ 5.4 GPa), which is consistent with previous report and demonstrates the emerge of superconductivity at high pressure [30]. The relationship between the resistivity anomaly and superconductivity as a function of pressure infers the thorough suppression of Lifshitz transition takes place at the onset of superconductivity. Similar phenomenon has been observed in electron-doped 122 iron-based superconductors, where the emergence of superconductivity coincides with the disappearance of the reconstructed Fermi surface rather than Fermi surface nesting [45]. Our results also establish the connection between superconductivity and Lifshitz transition in $Bi_4Br_4$. Moreover, since the theory predicts the non-trivial topology of $Bi_4Br_4$ even at high pressure [18], the possible coexistence of superconductivity and topological non-trivial feature engenders the exploration of Majorana fermions in this system.

To quantitively keep track on the temperature evolution of the energy gap, we plot the energy distribution curves (EDCs) at *M* point measured at different temperatures as a function of energy (FIG. 4(c)). The EDCs consist of the band gap between the signals from the LB and the UB. Although the positions of LB and UB show the strong temperature dependence, the gap value is almost unchanged with the varied temperature (FIG. 4(d)), excluding the possible band inversion and topological phase transition. Consequently, an energy gap around 0.24 eV of $Bi_4Br_4$ at room temperature promises its practical application to overcome the effect of thermal perturbation. Additionally, the temperature dependent shift of chemical potential leads to that the Fermi surface resides in the gap region of $Bi_4Br_4$ at room temperature, which provides the direct detection of its topological non-trivial edge states without the requirement of voltage gate to tune the Fermi level. Thus, the natural semiconducting behavior of $Bi_4Br_4$ at room temperature promotes the further research on its topological nature and potential applications on electronic and spintronic devices.

In summary, our ARPES spectra, STM data, and the first-principle calculation results match well with each other and provide the substantial evidence for the presence of the QSH insulating phase with a large energy gap in monolayer $Bi_4Br_4$. The STS curves reveal the 1D topological non-trivial edge state with a penetration depth around 1.7 nm away from the edges, which has been confirmed by first-principle calculations. The ARPES measurements demonstrate a temperature-induced Lifshitz transition of $Bi_4Br_4$ which is correlated to the resistivity anomaly in transport measurement. The emergence of superconductivity under high pressure coincides with the disappearance of the suppression of Lifshitz transition, which is analogous to that of the unconventional superconductor (pnitides). Our work not only creates the new platform to investigate the topological non-trivial nature of $Bi_4Br_4$ and promotes the accessible scope of this 1D material as building block to synthesize the practical nanoribbons and to engineer the new topological phase, but also provides the possibility of realizing topological superconductors with Majorana bound states.

The work was supported by the Beijing Natural Science Foundation (Z180007), the National Natural Science Foundation of China (11874003, 11904015, 12005251, 12004321, 12074021), UOW-BUAA Joint Research Centre, and the Natural Science Foundation of Hunan Province (No. 2019JJ50602).

Ming Yang, Chen Liu, and Yundan Liu contributed equally to this work.
[*] jincheng@buaa.edu.cn
[*] lijin@xtu.edu.cn
[*] cliu@ihep.ac.cn

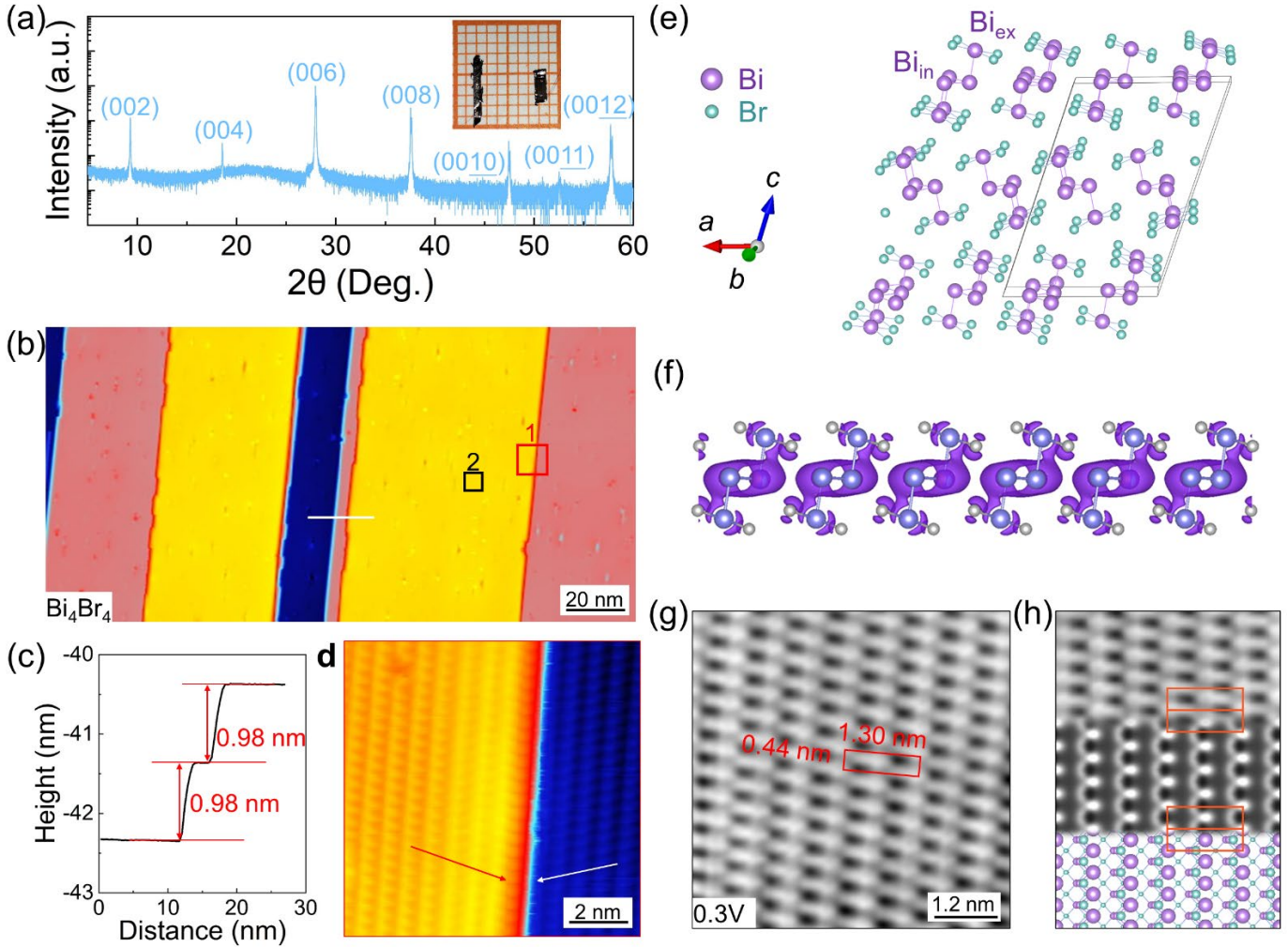

FIG. 1. Characterization of crystal structure of $Bi_4Br_4$. (a) X-ray diffraction measurement taken from the *ab* plane of α' $Bi_4Br_4$ crystal used in this work. Inset: A photography of $Bi_4Br_4$ crystals on the paper with millimeter grid. (b) Large-scale STM image of cleaved $Bi_4Br_4$ surface ($V_{bias}$ = 1 V, $I$ = 50 pA, 200 nm × 100 nm). (c) The line profile across the edge steps along the white solid line in (b). (d) High-resolution STM image of $Bi_4Br_4$ measured in area 1 in (b) labeled by red square ($V_{bias}$ = -0.4 V, $I$ = 50 pA, 10 nm × 10 nm). The red and white arrows are marked to show the mirror symmetry between adjacent layers. (e) Atomic structure of bulk $Bi_4Br_4$. The purple and blue balls denote the Bi atoms and Br atoms, respectively. (f) Atomic structure of monolayer $Bi_4Br_4$ and in-plane charge density difference. (g) High-resolution STM image of the terrace of $Bi_4Br_4$ measured in area 2 in (b) labeled by black square ($V_{bias}$ = 0.3 V, $I$ = 50 pA, 6 nm × 6 nm). (h) Schematic diagrams of the evolution from the optimized atomic structural models to simulated STM images, and then to the experimentally observed STM image (from bottom to up) for monolayer $Bi_4Br_4$.

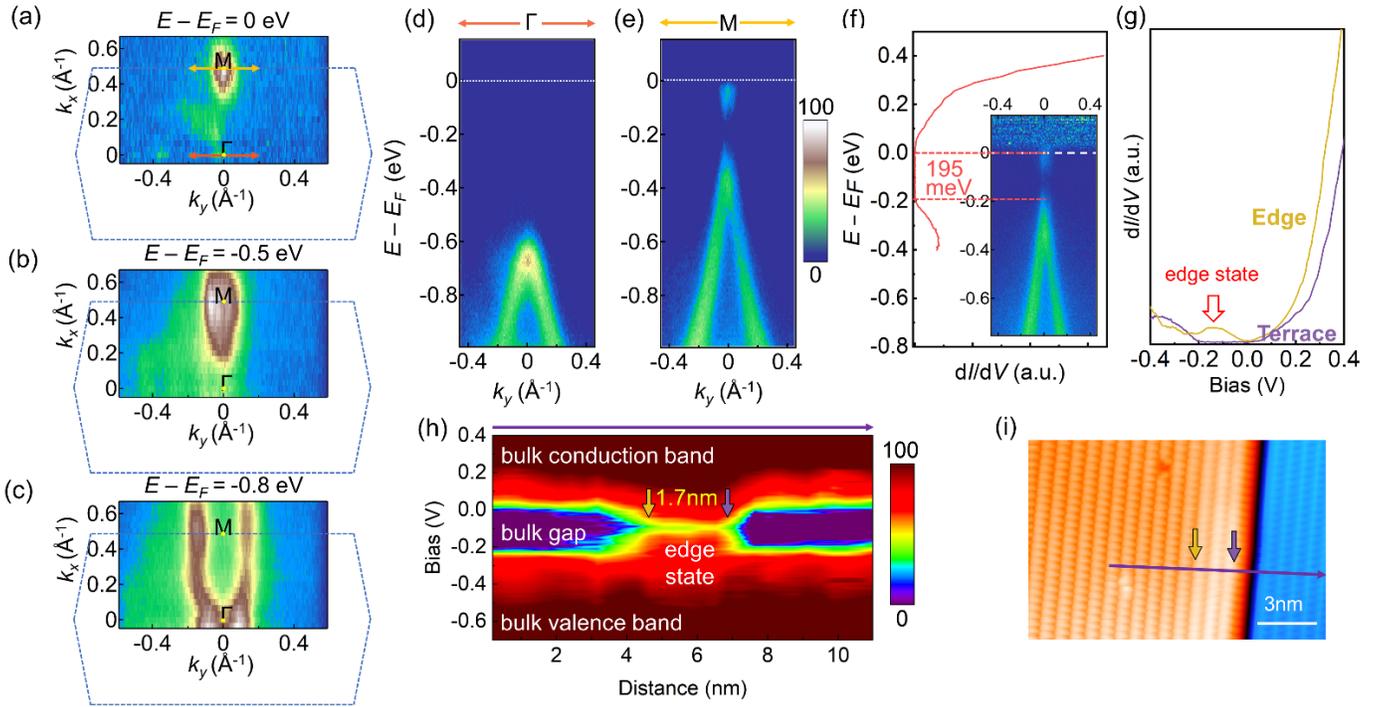

FIG. 2. ARPES and STM measurements of $Bi_4Br_4$. (a)-(c) The ARPES intensities of constant energy contours along a $k_y$-$k_x$ sheet plotted at the different energies: (a) $E = E_F$, (b) $E - E_F$ = -0.5 eV, and (c) $E - E_F$ = -0.8 eV. The blue dashed lines indicate the surface Brillouin zones of the (001) surface. The intensities are integrated within a ± 20 meV window about each binding energy. (d), (e) Energy-momentum dispersions in the 1D chain direction, measured along red mark and yellow mark, respectively. (f) STM d$I$/d$V$ spectrum acquired in the terrace of $Bi_4Br_4$ surface. The inset is the high-symmetry cut along the $\Gamma$-$M$-$\Gamma$ direction aligned in energy with the STS spectrum. (g) The d$I$/d$V$ spectra acquired in the edge (yellow) and inner terrace (purple) regions of $Bi_4Br_4$. The red and purple triangles are presented for the VBM and CBM, respectively. The red arrow is marked for the guidance of the position of edge state. (h) Spatially resolved d$I$/d$V$ data across the step edge in (i). The yellow arrow and purple arrow are marked for the guidance of the penetration depth of edge state. (i) STM topography covering the edge step between two adjacent layers ($V_{bias}$ = -0.5 V, $I$ = 50 pA, 15 nm × 10 nm). The purple line represents the locations of the measured STS curves.

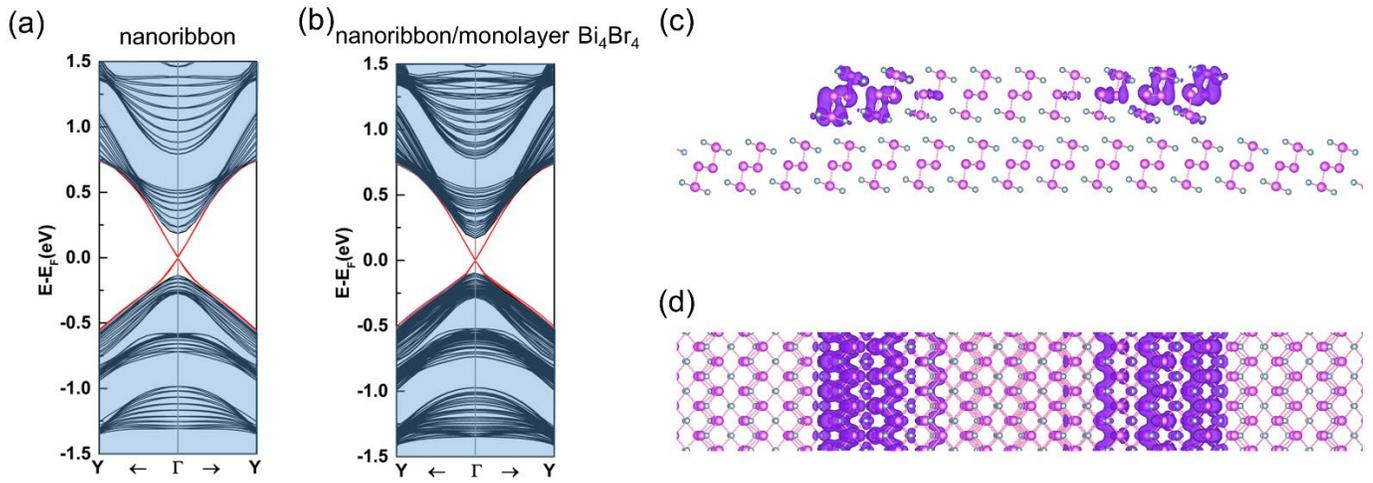

FIG. 3. Calculated topological non-trivial edge state of Bi$_4$Br$_4$ nanoribbon. (a) The electronic structure of monolayer Bi$_4$Br$_4$ nanoribbon with fully relaxed edges calculated with *ab initio* method. (b) Same as (a), but with a support with one monolayer Bi$_4$Br$_4$ substrate. The region colored in light-green in (a), (b) represents the states from the bulk, whereas the red curves are the states staying at the edges. (c), (d) The charge distribution of the edge state in real-space from both of (c) the side view and (d) the top view.

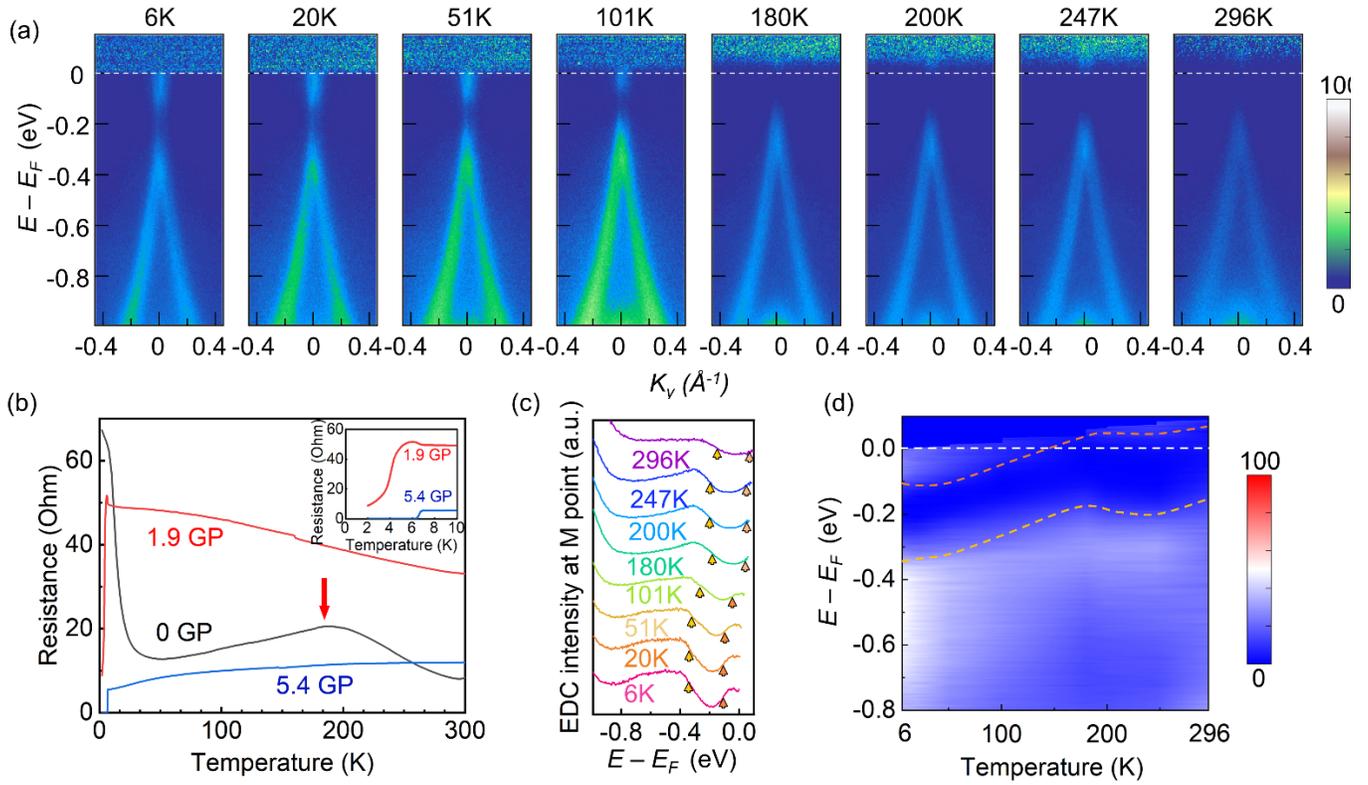

FIG. 4. Temperature-dependent ARPES measurements and the high-pressure transport results. (a) Temperature evolution of the band structures measured across $M$ point along the 1D chain direction. (b) Temperature-dependent resistance of a $Bi_4Br_4$ crystal under different pressures. Inset shows the transport results around superconducting transition temperature. (c) EDCs at $M$ point at different temperatures. The EDCs consist of the LB valence band and the UB conduction band with a valley separating in between them. The yellow and orange arrows are marked to the positions of LB maximum and UB minimum, respectively. (d) The spatially resolved EDCs as a function of energy and temperature. The yellow and orange dashed lines are labeled for the eye guidance of the gap.